\documentclass[12pt,preprint]{aastex}
\usepackage{amsmath}
\usepackage{amssymb}
\usepackage{xspace}
\usepackage{color}
\usepackage{ifthen}
\usepackage{xspace}

\newcommand{\XRBaiii}{zM}
\newcommand{\XRBaiv}{zm}
\newcommand{\XRBav}{ZM}
\newcommand{\XRBavi}{Zm}
\newcommand{\XRBa}[1]{%
\ifthenelse{#1 = 3}{\XRBaiii}{}%
\ifthenelse{#1 = 4}{\XRBaiv}{}%
\ifthenelse{#1 = 5}{\XRBav}{}%
\ifthenelse{#1 = 6}{\XRBavi}{}%
\xspace}

\newcommand{\powersep}{{\ensuremath{\times}}}

\newcommand{\g}{{\ensuremath{\mathrm{g}}}\xspace}

\newcommand{\cm}{{\ensuremath{\mathrm{cm}}}\xspace}
\newcommand{\yr}{{\ensuremath{\mathrm{yr}}}\xspace}
\newcommand{\km}{{\ensuremath{\mathrm{km}}}\xspace}
\newcommand{\Msun}{{\ensuremath{\mathrm{M}_{\odot}}}\xspace}
\newcommand{\Sec}{{\ensuremath{\mathrm{s}}}\xspace}
\newcommand{\erg}{{\ensuremath{\mathrm{erg}}}\xspace}
\newcommand{\ergs}{{\ensuremath{\erg\,\Sec^{-1}}}\xspace}

\newcommand{\kms}{{\ensuremath{\km\,\Sec^{-1}}}\xspace}

\newcommand{\gcc}{{\ensuremath{\g\,\cm^{-3}}}\xspace}

\newcommand{\Msunyr}{{\ensuremath{\Msun\,\yr^{-1}}}\xspace}

\newcommand{\lSect}[1]{{\label{sec:#1}}}

\newcommand{\Tabff}[1]{{\ref{tab:#1}}}
\newcommand{\Tab}[1]{{Table~\Tabff{#1}}}

\newcommand{\pan}[1]{{\textit{#1}}}

\newcommand{\FIGFF}[2]{{\ref{fig:#2}\pan{#1}}}

\newcommand{\FIG}[2]{{Fig.~\FIGFF{#1}{#2}}}
\newcommand{\Fig}[1]{{\FIG{}{#1}}}

\newcommand{\Sectff}[1]{{\ref{sec:#1}}}
\newcommand{\Sect}[1]{{\S~\Sectff{#1}}}

\newcommand{\Ep}[1]{{\ensuremath{10^{#1}}}}

\newcommand{\E}[1]{{\ensuremath{\powersep\Ep{#1}}}}

\shorttitle{RELATIVISTIC JETS FROM GRB PROGENITORS}
\shorttitle{Zhang, Woosley, \& Heger}

\begin{document}

\title{THE PROPAGATION AND ERUPTION OF RELATIVISTIC JETS FROM THE
STELLAR PROGENITORS OF GRBs} 

\author{Weiqun Zhang and S. E. Woosley}
\affil{
  Department of Astronomy and Astrophysics, 
  University of California,
  Santa Cruz, CA 95064
}
\email{zhang@ucolick.org, woosley@ucolick.org}

\and

\author{A. Heger}
\affil{
   Theoretical Astrophysics Group,
   T-6, MS B227,
   Los Alamos National Laboratory,
   Los Alamos, NM 87545
\\
   Enrico Fermi Institute,
   University of Chicago,
   Chicago, IL 60637 
}
\email{1@2sn.org}

\begin{abstract}
New two- and three-dimensional calculations are presented of
relativistic jet propagation and break out in massive Wolf-Rayet
stars.  Such jets are thought responsible for gamma-ray bursts.  As it
erupts, the highly relativistic jet core (3 to 5 degrees; $\Gamma
\gtrsim 100$) is surrounded by a cocoon of less energetic, but still
moderately relativistic ejecta ($\Gamma \sim 15$) that expands and
becomes visible at larger polar angles ($\sim$ 10 degrees). These less
energetic ejecta may be the origin of X-ray flashes and other
high-energy transients which will be visible to a larger fraction of
the sky, albeit to a shorter distance than common gamma-ray
bursts. Jet stability is also examined in three-dimensional
calculations. If the jet changes angle by more than three degrees in
several seconds, it will dissipate, producing a broad beam with
inadequate Lorentz factor to make a common gamma-ray burst.
This may be an alternate way to make X-ray flashes.
\end{abstract}

\keywords{gamma rays: bursts --- hydrodynamics --- methods: numerical --- relativity}

\section{INTRODUCTION}
\lSect{intro}

It is now generally acknowledged that ``long-soft'' gamma-ray bursts
(henceforth, just ``GRBs'') are a phenomenon associated with the
deaths of massive stars. In addition to the observed association with
star-forming regions in galaxies \citep{vre01,blo02a,gor03}; ``bumps''
observed in the afterglows of many GRBs
\citep{rei99,gal00,blo02b,gar03,blo03}; spectral features like a
WR-star in the afterglow of GRB 021004 \citep{mir02}; and the
association of GRB 980425 with SN 1998bw \citep{gal98}, there is now
incontrovertible evidence that at least one GRB (030329) was
accompanied by a bright, energetic supernova of Type Ic, SN 2003dh
\citep{pri03,hjo03,sta03}.  Thus some, if not all GRBs, are produced
when the iron core of a massive star collapses either to a black hole
\citep{woo93,Mac99} or a very rapidly rotating highly magnetic neutron
star \citep{whe00}, producing a relativistic jet. Other alternatives
in which the GRB occurs {\sl after} the supernova \citep{vie98} cannot
simultaneously explain GRB 030329 and SN 2003dh.

At the same time, the general class of high energy transients once
generically called ``gamma-ray bursts'' has been diversifying. For
some time, soft gamma-ray repeaters (SGRs) and ``short hard gamma-ray
bursts'' have enjoyed a distinct status and, presumably, a separate
origin.  In addition, we now have cosmological ``X-ray flashes''
\citep[``XRFs'';][]{Hei01,Kip03}, ``long, faint gamma-ray bursts''
\citep{zan03}, and lower energy events like GRB980425
\citep{gal98}. Is a different model required for each new phenomenon,
or is some unified model at work, as in active galactic nuclei, a
model whose observable properties vary with its environment, the angle
at which it is viewed, and perhaps its redshift?

The answer is probably ``both''. Even within the confines of the
collapsar model, there are Types I, II and III \citep{Heg03b}. Type I
happens only in massive stars that make their black holes promptly
\citep{Mac99}. This is what people generally have in mind when they
use the word ``collapsar''. In Type II though, a similar mass black
hole forms by fall back and the burst lasts much longer
\citep{mac01}. Type III, happens only at very low metallicity and
requires Pop III stars as its progenitors \citep{fry01}. The event is
very energetic, but highly redshifted. These other models may be
particularly appropriate for unusually long bursts. A jet that wavered
in its orientation in any of these events would emerge with a greater
load of baryons. So would a jet whose power declined in a time less
than that required to traverse the star.  A jet in a blue supergiant
would become choked because the power of the central engine has
decreased substantially when the jet is still deep inside the star.
Thus any sort of collapsar happening in a blue supergiant instead of a
Wolf-Rayet star would produce a different sort of transient.  Finally,
it may just be that jets are just born with different baryon loadings,
perhaps relating to the mass of the presupernova star and its
distribution of angular momentum and magnetic field.  Our present
understanding of nature allows for all these solutions.

However, it also seems to reason that not all jets are the same, even
for Type I collapsars, and that even if they were, different phenomena
would be seen at different angles.  We are thus motivated to consider
the observational consequences of highly relativistic jets as they
propagate through, and emerge from massive stars. What would they look
like if seen from different angles? In particular, what is the
distribution with polar angle of the energy and Lorentz factor?
Might softer, less energetic phenomena be observed off axis and with
what frequency?

Jets inside massive stars have been studied numerically in both
Newtonian \citep{Mac99, mac01,Kho99} and relativistic simulations
\citep[][henceforth Paper 1]{alo00, zha03} and it has been shown that
the collapsar model is able to explain many of the observed
characteristics of GRBs.  These previous studies have also raised
issues which require further examination, especially with higher
resolution.  For instance, the emergence of the jet and its
interaction with the material at the stellar surface and the stellar
wind could lead to some sort of ``precursor'' activity.  The cocoon of
the jet will also have different properties from the jet itself and
shocks within the cocoon or with external matter could lead to
$\gamma$-ray and hard X-ray transients \citep{ram02}. There is also
the overarching question of whether jets, calculated in two dimensions
with assumed axial symmetry, are stable when studied in three
dimensions.

Here we examine in two- and three-dimensional numerical studies, the
interaction of relativistic jets with the outer layers of the
Wolf-Rayet stars thought responsible for GRBs. We follow the emergence
of the jets for a sufficient length of time to ascertain the
properties of the cocoon explosion that surrounds the GRB and we study
the dependence of the results on the dimensionality of the
calculation. Such explosions are visible to a much greater angle and
to a much larger number of observers. The observed phenomenon could be
a hard X-ray flash or a weak GRB.  We also find that the properties of
the emergent jet are quite sensitive to whether the jet maintains its
orientation (to within a critical angle) for durations of $10\,\Sec$
or so, the time it takes the jet to traverse the star. Jets that
waiver may make hard transients of some sort, but not GRBs.

\section{Stellar Model and Numerical Methods}

\subsection{Progenitor Star}

A collapsar is formed when the iron core of a rotating massive star
collapses to a black hole and an accretion disk. The interaction of
this disk with the hole, through processes that are still poorly
understood, produces jets with a high energy to mass ratio. Here we
are concerned primarily with the propagation of these relativistic
jets, their interactions with the star and the stellar wind, and the
observational implications, and not so much with how they are born. In
fact, our results here would be the same if the jet were produced by
some other means \citep{whe00} inside the same star.

\begin{figure}
\includegraphics{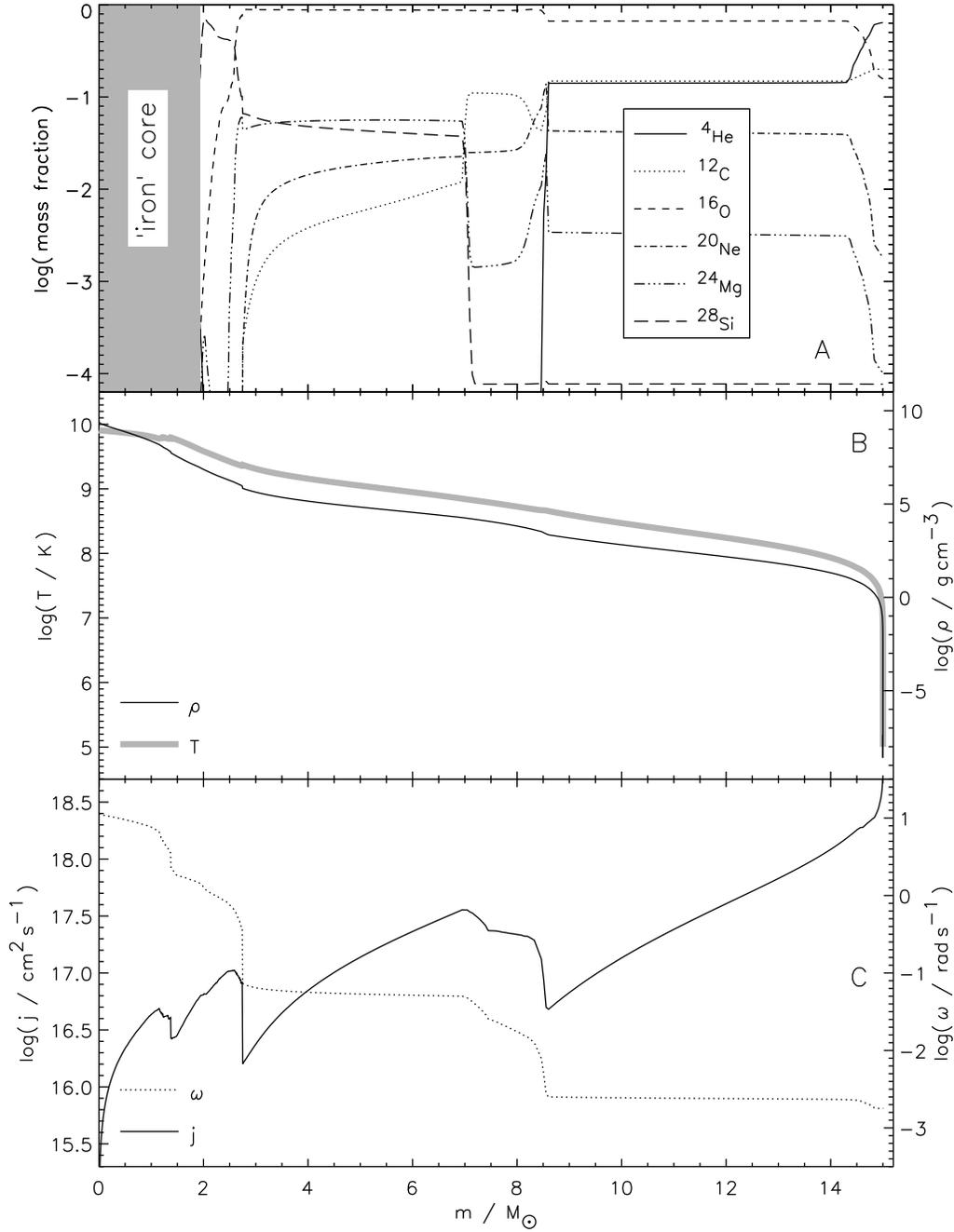}
\caption{Structure of the star at the onset of core collapse as a
function of enclosed mass, $m$.  Panel~A gives mass fractions of the
dominant species, Panel~B density ($\rho$) and temperature ($T$)
stratification, and Panel~C angular velocity ($\omega$) and average
specific angular momentum on spherical shells ($j$).  The enclosed
mass at $10^{10}\,\cm$, where the jets were launched in our
simulations, is $10.3\,\Msun$.
\label{fig:prog}}
\end{figure}

Unlike in our previous study (Paper 1), the initial stellar model is
taken directly from a presupernova star with a very finely zoned
surface (Table~\ref{tab:proj}; Fig.~\ref{fig:prog}).  Because we are
only interested in events that happen outside about $10^{10}\,\cm$ on
a time scale of, at most, tens of seconds, modification of the initial
star by core collapse and rotation is negligible. Specifically, our
initial model is a $15\,\Msun$ helium star calculated by
\citet{heg03}.  This helium star has been evolved to iron-core
collapse while following the transport of angular momentum and
including the effects of rotational mixing as in \citet{Heg00}.  The
initial star was a nearly pure helium star rotating rigidly with a
surface velocity equivalent to 10\% Keplerian. This is a typical value
in the study of \citet{Heg00}.  Presumably the star lost its envelope
to a companion early in helium burning.  Further mass loss and the
transport of angular momentum by magnetic fields were ignored so as to
form a big iron core that would very likely collapse to a black hole
and give ample angular momentum for a disk.  The radius of the helium
star at onset of collapse is $8.8 \times 10^{10}\,\cm$ and the surface
of the star is very finely zoned (surface zoning $\lesssim
10^{21}\,\g$). Despite the fact that angular momentum transport was
followed in the initial model to show its suitability as a collapsar
(\Fig{prog}), rotation plays no role in the present study.

Outside the star, the background density, which comes from the stellar
wind, is assumed to scale as $R^{-2}$. The actual value of this
density, unless it is very high, plays no role in determining the
final answer at the small scales considered here ($<2 \times 10^{12}$
cm), but finite value is needed in order to stabilize the code. We
assumed a value at $R = 10^{11}\,\cm$ of $5 \times 10^{-11}\,\gcc$,
where $R$ is the distance from the center of the star.  This
corresponds to a mass loss rate of $\sim 1\times10^{-5}\,\Msunyr$ for
a wind velocity of $\sim 1000\,\kms$ at $10^{11}\,\cm$.  This mass
loss rate is rather low for typical Wolf-Rayet stars of this mass in
our galaxy \citep{Nug00}, but might be more appropriate for the low
metallicity in the GRB neighborhood \citep{Cro02}.  A small value was
also taken in order to be consistent with the assumed progenitor
structure that was calculated with zero mass loss.

\subsection{Computer Code}

We employed a multi-dimensional relativistic hydrodynamics code that
has been used previously to study relativistic jets in the collapsar
environment (Paper 1).  Briefly, our code employs an explicit Eulerian
Godunov-type shock-capturing method \citep{alo99}.  The governing
equations of relativistic hydrodynamics with a causal equation of
state can be written as a system of conservation laws for rest mass,
momentum, and energy.  To solve the equations numerically, each
spatial dimension is discretized into cells.  Using the method of
lines, the multi-dimensional time-dependent equations can then be
evolved by solving the fluxes at each cell interface.  In order to
achieve high order accuracy in time, the time integration is done
using a high order Runge-Kutta scheme \citep{shu88}, which involves
prediction and correction.  An approximate Riemann solver
\citep{alo99} using the Marquina's algorithm is used to compute the
numerical fluxes from physical variables: pressure, rest mass density,
and velocity at the cell interface.  The values of the physical fluid
variables at the cell interface are interpolated using reconstruction
schemes \citep[e.g., piecewise parabolic method,][]{col84}.  This
reconstruction procedure ensures high order accuracy in spatial
dimensions.  The conserved variables: rest mass, momentum and energy
are evolved directly by the scheme.  In each time step, physical
variables, such as pressure, rest mass density and velocity, which are
necessary in calculating the numerical fluxes, can be recovered from
conserved variables by Newton-Raphson iteration.  The code can operate
in Cartesian, cylindrical, or spherical coordinates.  Approximate
Newtonian gravity is implemented by including source terms in the
equations.  Total energy, which includes kinetic and internal energy,
in the laboratory frame are not exactly conserved due to gravitational
potential energy and the way of implementing gravity as source terms.
However, the code conserves energy to machine precision when gravity
is turned off.  Furthermore, the gravitational potential energy is
negligible for relativistic material in our calculations.  For instance,
assuming a $10\,\Msun$ central point mass, a fluid element with a
velocity of $0.5\,c$ at the inner boundary of our computational grids,
$10^{10}\,\cm$ (\Sect{2dsetup}, \Sect{3ddef}), has potential energy of
less than 0.1\% of its kinetic energy.  Thus we expect that our
calculations conserve energy to an accuracy of better than 0.1\%.  In
simulations present in this paper, A gamma-law equation of state with
$\gamma = 4/3$ is used for the simulations presented in this paper.

\section{Two-Dimensional Models}
\lSect{2d}

\subsection{Model Set Up and Definitions}
\lSect{2dsetup}

The mass interior to $1.0 \times 10^{10}\,\cm$ is removed from the
presupernova star and replaced by a point mass.  No self gravity is
included. This should suffice since we are studying phenomena that
happen on a relativistic time scale and the speed of sound is very
sub-luminal. While jets presumably go out both axes, we follow here
only one, assuming symmetry in the other hemisphere. Jets are injected
along the rotation axis (the center of the cylindrical axis of the
grid) through the inner boundary.  Each jet is defined by its power
(excluding rest mass energy), $\dot{E}$, its initial Lorentz factor,
$\Gamma_0$, and the ratio of its total energy (excluding rest mass
energy) to its kinetic energy, $f_0$.  Our previous studies in Paper 1
have shown that a jet that starting with a half-opening angle of
$20\degr$ and a high Lorentz factor, $\Gamma \sim 50$ at $2000\,\km$,
will be shocked deep inside of the star. By the time it reaches
$10^{10}\,\cm$, a jet should have a large ratio of internal energy to
rest mass, a half-opening angle of about $5\degr$, and a Lorentz
factor, $\Gamma \sim 5 - 10$.

Though GRBs observationally have highly variable properties, the jet
power was taken here to be constant for the first $20\,\Sec$, then
turned down linearly during the next 10 seconds. That is, during the
interval 20 to 30 s, the power scaled as $(30\,\Sec - t)/10\,\Sec$
decreasing to zero at $30\,\Sec$.  During the declining phase, the
pressure and density remained constant, and the Lorentz factor was
calculated from the internal energy, density, and power.  After 30
seconds, a pure outflow boundary condition was used for the inner
boundary.  The axis of the jet is defined as the $z$-axis in all cases
and the jets were initiated parallel to that axis in a region that
subtended a half-angle of 5 degrees as viewed from the origin.

Four models were calculated which span a range of energies and Lorentz
factor (Table~\ref{tab:mod2}).  In Model 2A, a total energy of $\sim 5
\times 10^{51}\,\erg$ is injected, comparable to the results of
\citet{fra01}, but larger than the results of \citet{pan01} by an
order of magnitude (Table~\ref{tab:energy}).  In Models 2B and 2C,
higher and lower energy deposition rates were employed. For the
parameters chosen, a jet with an initial Lorentz factor of 5 or 10 at
$10^{10}\,\cm$ should have a final Lorentz factor of $\sim 180-200$ if
all internal energy is converted into kinetic energy.  The zoning
employed in the models is given in \Tab{2dgrid}. Typically over three
million zones were used.

Model 2C used an extended grid in the $r$-direction, which allowed the
greater lateral expansion of the jet to be followed at late times.
This was necessary because of its lower energy and greater expansion
during the time it took to traverse the extent of the $z$-grid.  Model
2T used conditions like those of Model 2B, but a grid that was both
smaller and coarser, chosen to be equivalent to that used in the
three-dimensional calculations described later in the paper
(\Sect{3d}).  

\subsection{Results in Two Dimensions}
\lSect{2dresults}

\begin{figure}
\epsscale{1.0}
\plotone{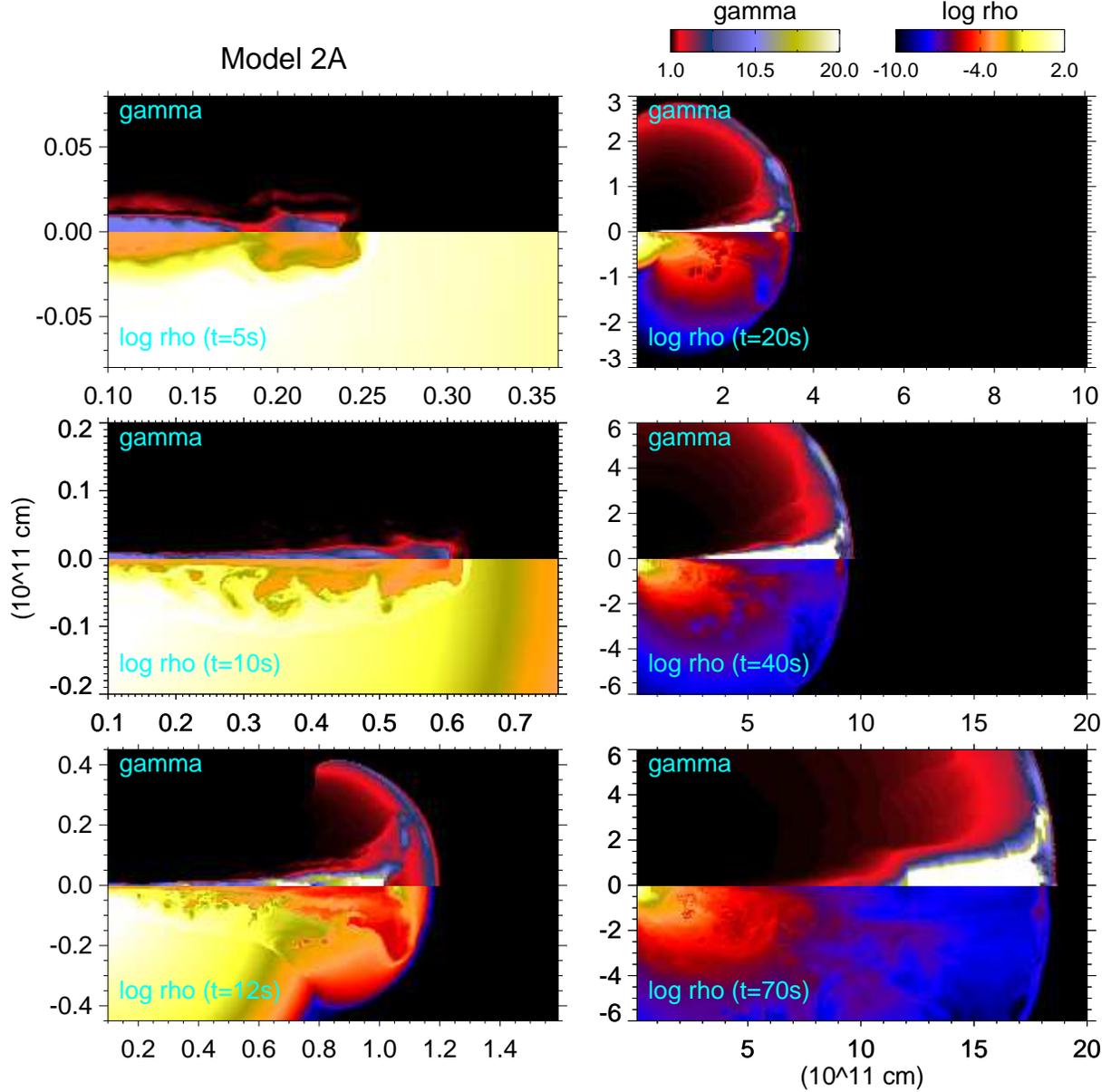}
\caption{Time evolution of the density (bottom of each frame) and
Lorentz factor (top of each frame) in the jet and its environs for
Model 2A. The density is on a logarithmic scale, the Lorentz factor is
on a linear scale, both color coded. Quantities are given 5, 10, 12,
20, 40, and 70 s after the initiation of the jet at $0.1 \times
10^{11}\,\cm$. The third panel at 12\,s is just as the jet is erupting
from the star (radius $0.89 \times 10^{11}\,\cm$). By 70\,s, most of the
internal energy has converted to kinetic and the Lorentz factor has
almost reached its terminal value.  Values of $\Gamma$ at this time in
the central, over-exposed jet are near 150. Note the explosion of the
cocoon as the jet erupts and the ejection of material with terminal
Lorentz factor, $\Gamma \sim 10 - 20$, at angles around 10 degrees off
axis.
\label{fig:2A}}
\end{figure}

\begin{figure}
\epsscale{1.0}
\plotone{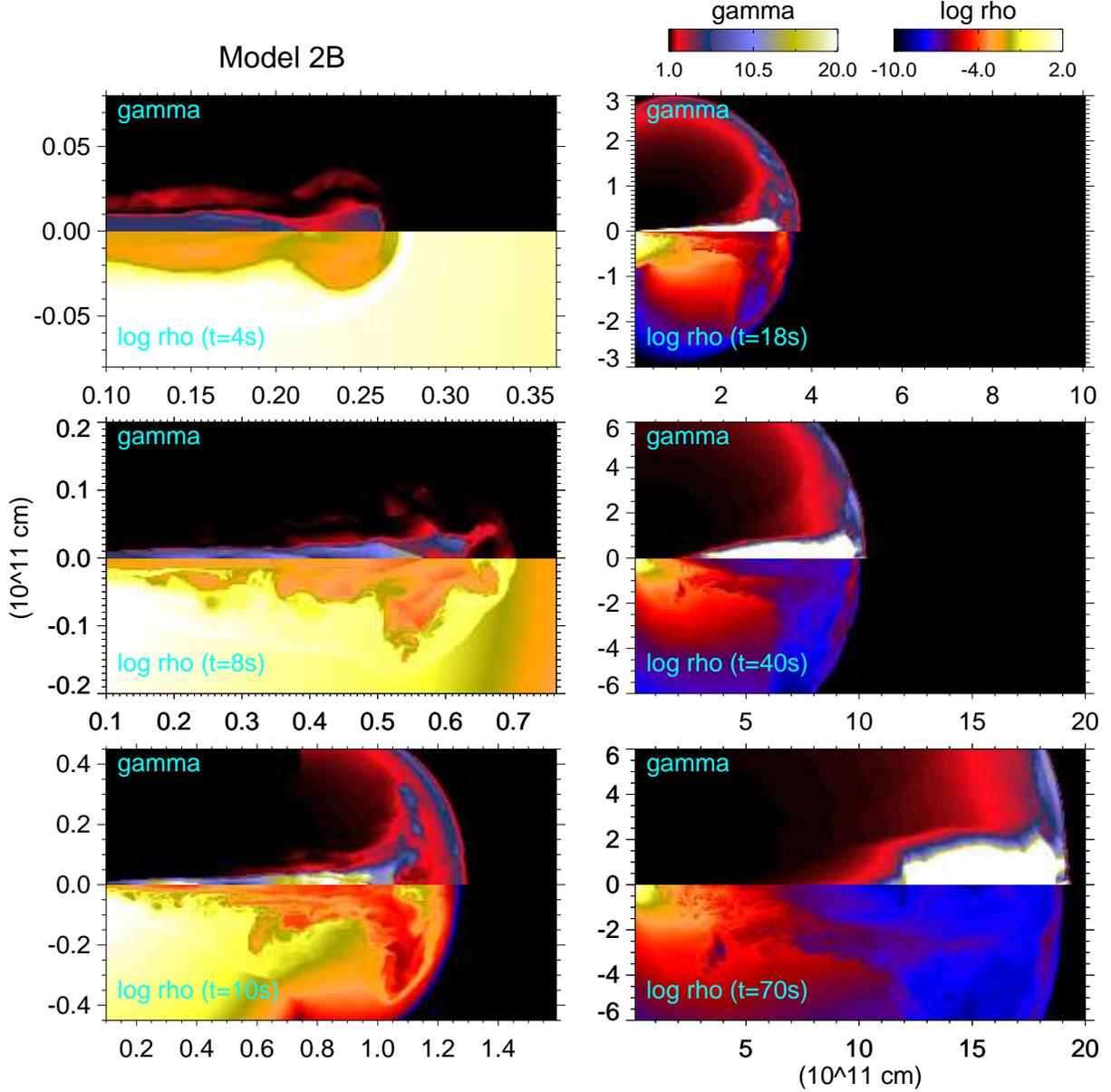}
\caption{Time evolution of the density (bottom of each frame) and
Lorentz factor (top of each frame) in the jet and its environs for
Model 2B. The density is on a logarithmic scale, the Lorentz factor is
on a linear scale, both color coded. Quantities are given 4, 8, 10,
18, 40, and 70 s after the initiation of the jet at $0.1 \times
10^{11}\,\cm$. See also Fig.~\ref{fig:2A}.  Model 2B is a more
energetic jet and reaches the surface in a shorter time.
\label{fig:2B}}
\end{figure}

\begin{figure}
\epsscale{1.0}
\plotone{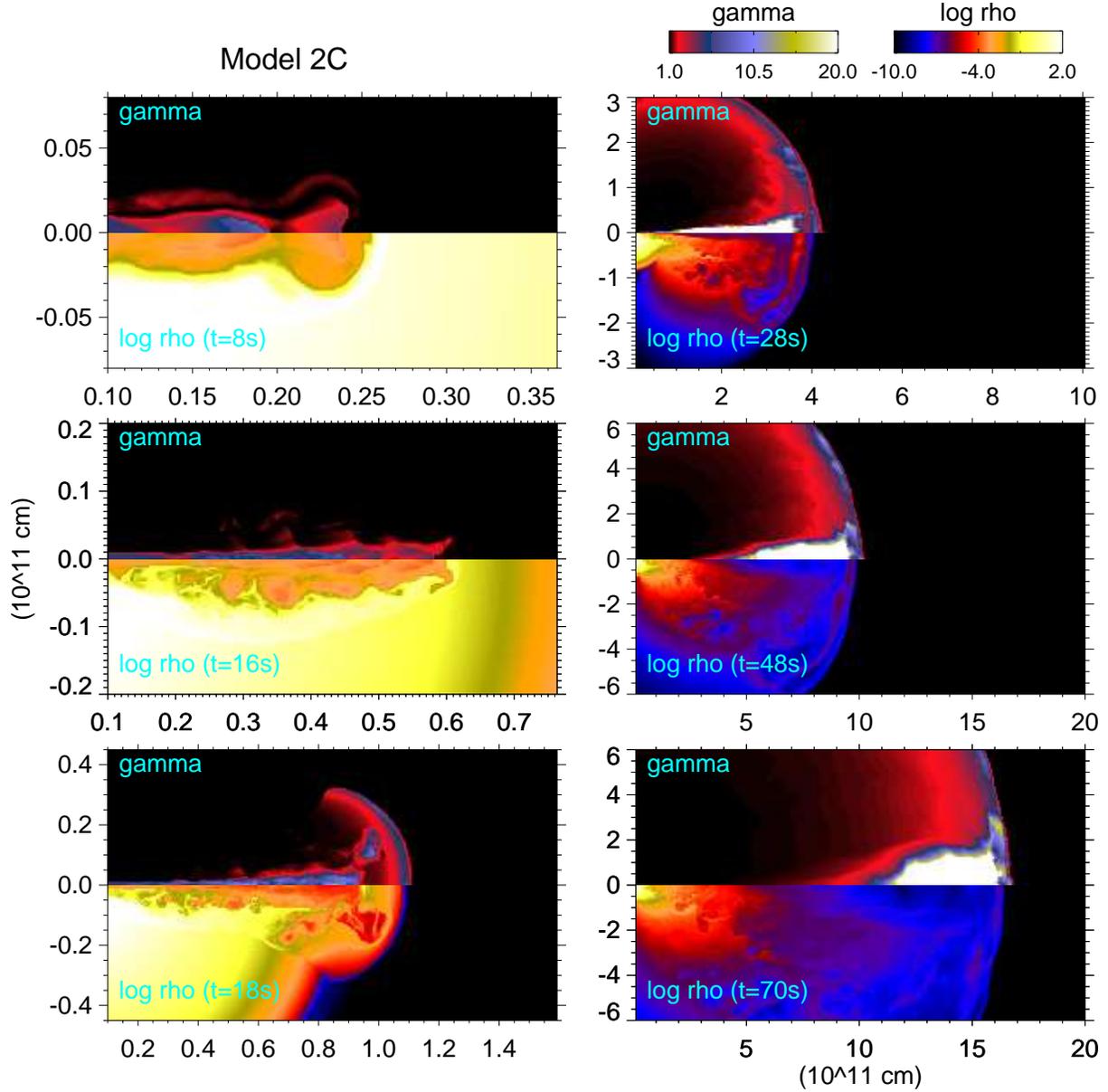}
\caption{Time evolution of the density (bottom of each frame) and
Lorentz factor (top of each frame) in the jet and its environs for
Model 2C. The density is on a logarithmic scale, the Lorentz factor is
on a linear scale, both color coded. Quantities are given 8, 16, 18,
28, 48, and 70 s after the initiation of the jet at $0.1 \times
10^{11}\,\cm$. See also Fig.~\ref{fig:2A}.  Model 2C is a less
energetic jet and takes longer to reach the surface.
\label{fig:2C}}
\end{figure}

The relativistic jet begins to propagate along the $z$-axis shortly
after its initiation.  In agreement with previous studies \citep[][and
Paper 1]{alo00}, the jet consists of a supersonic beam, a shocked
cocoon, a bow shock, and is narrowly collimated. Some snapshots of
Models 2A, 2B, and 2C are given in Figs.~\ref{fig:2A}, \ref{fig:2B},
and \ref{fig:2C}. The resulting ``equivalent isotropic energies'' for
matter exceeding a certain Lorentz factor are also given, at
$70\,\Sec$, for the same three models in Figs.~\ref{fig:eiso2a},
\ref{fig:eiso2b}, and \ref{fig:eiso2c}. The latter figures also give
the estimated terminal Lorentz factor assuming that internal energy
along the radial line of sight is converted into kinetic energy. The
equivalent isotropic energy is defined as that energy an isotropic
explosion would need in order to give the calculated flux of energy
along the line of sight. It is a way of mapping the highly asymmetric
two-dimensional results into equivalent one-dimensional models.  The
change in Lorentz factor between $70\,\Sec$ and infinity is not very
great, except in situations where matter that was highly relativistic
to begin with receives an additional boost in its frame by expansion.
The duration of the calculation was set by how long it took the jet to
reach the end of the simulation grid and would be costly to increase,
but as the figures show, especially in Model 2C, there was still an
interesting amount of internal energy at $70\,\Sec$.

\begin{figure}  
\epsscale{0.9}
\plotone{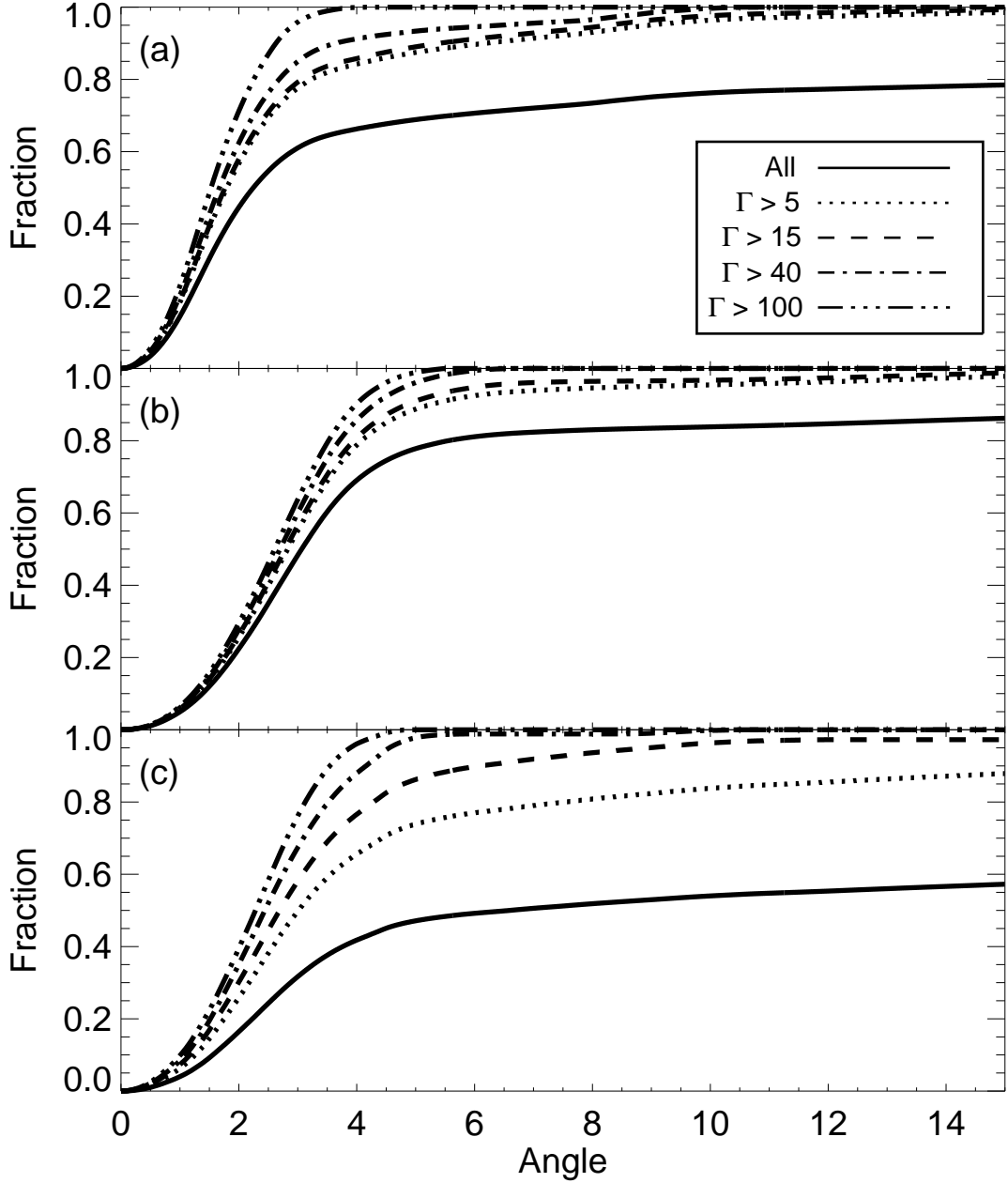}
\caption{Fraction of energy inside a certain angle to the total energy
on the grid for (a) Model 2A, (b) Model 2B, and (c) Model 2C.
Different lines are for material with different Lorentz factors.  It
is clearly shown that highly relativistic material is confined to a
small angle $\sim 4\degr$.  For Model 2C, mildly relativistic ($\Gamma
> 5$) material at larger angles ($\theta > 10$) contains about $20\%$
of its total in mildly relativistic energy.  This fraction is smaller for
Models 2A and 2B.
\label{fig:fraction}}
\end{figure}

The fractions of energies inside a certain angle to the total energies
on the grid are given in Fig.~\ref{fig:fraction}.  In all cases
studied the high Lorentz factor characteristic of common GRBs is
confined to a narrow angle of about 3 to 5 degrees with a maximum
equivalent isotropic energy in highly relativistic matter along the
axis of $\sim 3 \times 10^{53}$ to $3 \times 10^{54}\,\erg$. At larger
angles there is significant energy, though, and Lorentz factor,
$\Gamma \sim$ 10 to 20. At an angle of 10 degrees for example, the
equivalent isotropic energy in matter with $\Gamma > 20$ is $\sim
10^{52}\,\erg$ in Model 2A and even $10^{51}\,\erg$ in Model 2C. At
larger angles there is less energy, but still the possibility of low
power transients of hard radiation.  Model 2B did not eject much
material with $\Gamma \gtrsim 10$ at angles $\sim 10$ degrees.  The
equivalent isotropic energy at larger angles ($ > 2\degr$) for all
three models can be fit well by a simple power-law, $E_0 \times
(\theta/2\degr)^{-3}\,\erg$, where $E_0$ is $1.5 \times 10^{54}$, $4.5
\times 10^{54}$ and $6.8 \times 10^{53}$, for models 2A,
2B and 2C, respectively.  
The ratios of the values of $E_0$, $1.5 : 4.5 : 0.68$, are very close
to those of the energy deposition rates, $1.0 : 3.0 : 0.5$.
Inside $2\degr$, the distributions of energy
and Lorentz factor are roughly flat, $\sim E_0$.  The values of $E_0$
can be roughly estimated from the total injected energy and the percentage
of the energy in the jet core.  For example, about $40\%$ of
the total energy on the grid is contained inside 2 degrees for
Model~2A (Fig.~\ref{fig:fraction}).  About $5 \times 10^{51}\,\erg$
(Table~\ref{tab:mod2}) was injected for both jets in Model~2A.  One
can estimate that the equivalent isotropic energy inside 2 degrees is
$\sim 1.6 \times 10^{54}\,\erg$ for Model~2A.  More models need to be
calculated to find how the distributions of energy and Lorentz factor
at breakout depend on initial conditions.  

\Tab{energy} gives the energies for various components of the ejecta
at a time $40\,\Sec$ after the jet was initiated. There still remains
considerable internal energy available for conversion to kinetic at
this time (which was chosen as a time when most of the relativistic
matter was still on the grid), but this will affect mostly the matter
with high Lorentz factor, $\Gamma > 10$. The results show that most of
the energy injected as jets at $10^{10}\,\cm$ emerges in matter that
is still relativistic. Only a small fraction goes into
sub-relativistic expansion, nominally $v < 0.5\,c$ and consequently,
only a little of the jet energy is used to blow up the star. If there
were no other energy sources the kinetic energy of the supernova would
be $\lesssim 2 \times 10^{51}$ erg.

The total energy of relativistic ejecta is also useful for comparison
to radio observations of the afterglows of GRBs. For example,
\citet{Li99} have placed limits on the total energy of ejecta with $v
> 0.5\,c$ in SN 1998bw. The limit, $3 \times 10^{50}\,\erg$, may be an
approximate estimate of the actual energy. The corresponding energy
for Model 2C here is four times larger (Table~\ref{tab:energy}).  It
seems likely that lower energy models than Model 2C could be
constructed that would still give high Lorentz factors and equivalent
isotropic energies in a narrow range of angles around the polar axis.
That is, GRB 980425 may have been a harder, more energetic GRB seen off
axis. However, the low energy of that burst, some five orders of
magnitude less that that expected for a centrally observed GRB
(Fig.~\ref{fig:2C}) remains surprising. Either GRB 980425 was observed
at a polar angle larger than 15 degrees, or the burst was weaker at
all angles because of baryon loading (\Sect{intro}; Woosley, Eastman,
\& Schmidt 1999).

The possibility that the off-axis emission from material with $\Gamma
\sim$ 10 to 20 corresponds to XRFs is discussed in \Sect{disc}.

\begin{figure}  
\plotone{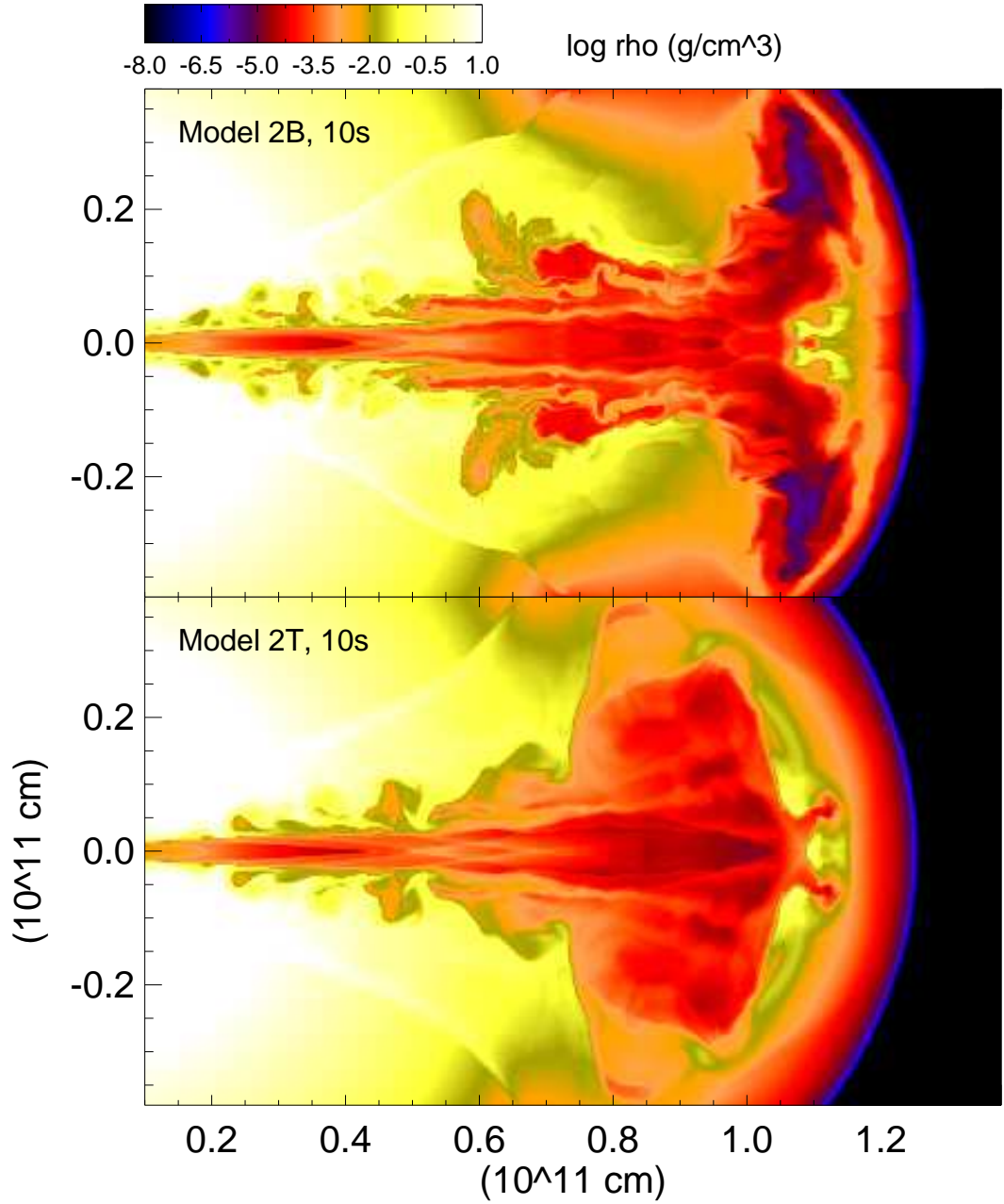}
\caption{Resolution study in two dimensions. The jets in Models 2B and
2T had identical parameters, but the calculation was carried out in
cylindrical grids having different resolution (\Tab{2dgrid}). Model 2T
had the lower resolution. \label{fig:2bt}}
\end{figure}

\begin{figure}  
\plotone{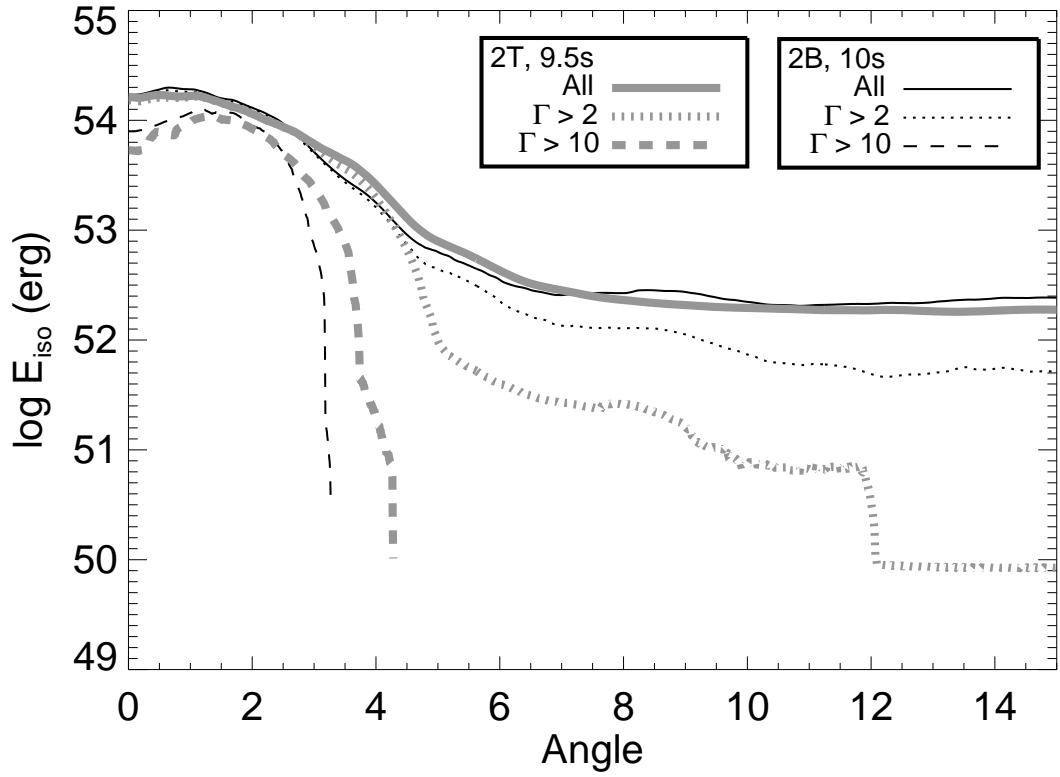}
\caption{A comparison of equivalent isotropic energy for Models 2B and
2T.  The thin and dark lines show the equivalent energy for Model 2B
at $10\,\Sec$; the thick and light lines show the equivalent energy
for Model 2T at $9.5\,\Sec$.  See also Tables~\ref{tab:mod2},
\ref{tab:2dgrid} and Fig.~\ref{fig:2bt}. \label{fig:eiso2bt}}
\end{figure}

In 2D simulations with cylindrical coordinates, there is an imposed
symmetry axis of the coordinate system.  It is very important to
repeat these calculations in 3D to ensure that our 2D results are
valid and to examine three-dimensional jet instabilities.
Three-dimensional calculations are, however, computationally
expensive.  So we have to use lower resolution for 3D calculations.
In order to compare 2D and 3D results with the same resolution and
identical initial parameters, we did a lower resolution 2D run, Model
2T (Tables~\ref{tab:mod2} and \ref{tab:2dgrid}).  A comparison of two-
and three-dimensional results will be discussed in \Sect{3d}.  It is
also interesting to compare Model 2B and 2T, which had identical
parameters but in different resolution.  Qualitatively the results are
similar.  In both models, the jet emerges from the star with a cocoon
surrounding the jet beam and a dense ``plug'' at the head of the jet
(Fig.~\ref{fig:2bt}).  There are, however, noticeable differences, as
we expected.  In particular, it takes less time for the jet in Model
2T to emerge from the star than that in Model 2B, because they have
different numerical viscosity.  The larger numerical viscosity in
Model 2T is due to its lower resolution.  Despite morphologic
differences between Model~2B and 2T (Fig.~\ref{fig:2bt}), the
energetics of these two models are quite similar (Table~\ref{tab:2bt},
Fig.~\ref{fig:eiso2bt}).  The distributions of equivalent isotropic
energy versus angle for the jet core ($\lesssim 3$ degrees) in Models
2B and 2T are very similar.  However, Model 2B has more mildly
relativistic material ($2< \Gamma <10$) in its cocoon than Model 2T.

\section{Three-Dimensional Models}
\lSect{3d}

For the three-dimensional models, the same helium star was remapped
into a three-dimensional Cartesian grid.  Because of the greater cost
of computations, these studies covered only an interval of $10\,\Sec$,
adequate to watch the jet propagate, develop a cocoon and break out,
but not long enough to study the the expansion after break out to any
great extent.

\subsection{Model Definitions}
\lSect{3ddef}

The parameters of the jet, its power, Lorentz factor and energy
loading, were all identical to those of Model 2B, that is $\Gamma =
5$, $\dot{E} = 3 \times 10^{50}\,\ergs$ and $f_o = 40$ (\Tab{mod3}).
The grid employed in all cases was Cartesian with 256 zones each along
the $x$- and $y$-axes and 512 along the $z$-axis (jet axis)
(\Tab{3dgrid}).

The initial model and the initiation of the jet in Model 3A were
perfectly symmetric with regard to the axis of the jet.  The perfect
symmetry was maintained in the calculation because our numerical
scheme did not break any symmetry.  In order to break the perfect
symmetry of the cylindrical initial conditions, an asymmetric jet was
injected in Models 3BS and 3BL.  At the base of the jet in Model 3BS,
pressure and density were 1\% more than those of Model 3A if $y \ge
{\tan{\alpha}}\,x$, and 1\% less otherwise, where $\alpha = 40$
degrees.  Hence the jet in Model 3BS has a $\pm 1\%$ imbalance in
power.  In Model 3BL, a $\pm 10\%$ imbalance in power was employed.

Several other models were calculated to explore the collimation
properties of non-radial jets that precessed. These jets were
initiated at the same $10^{10}\,\cm$ inner boundary with a half-angle
of 5 degrees as measured from the center. However, they were given a
non-$z$ (symmetry axis) component of momentum that would have
resulted, in a vacuum, in a propagation vector inclined to the
$z$-axis by 3 degrees (Model 3P3), 5 degrees (Model 3P5), and 10
degrees (Model 3P10). The jet precessed around the $z$-axis with a
period, in all cases of $2\,\Sec$.  This period was chosen to be short
compared with the time it took the jet to traverse the remaining star
between $10^{10}\,\cm$ and the surface, yet long enough that the jet
would still be distinct from a cone.

\subsection{Breakout in Three Dimensions}

\begin{figure}  
\plotone{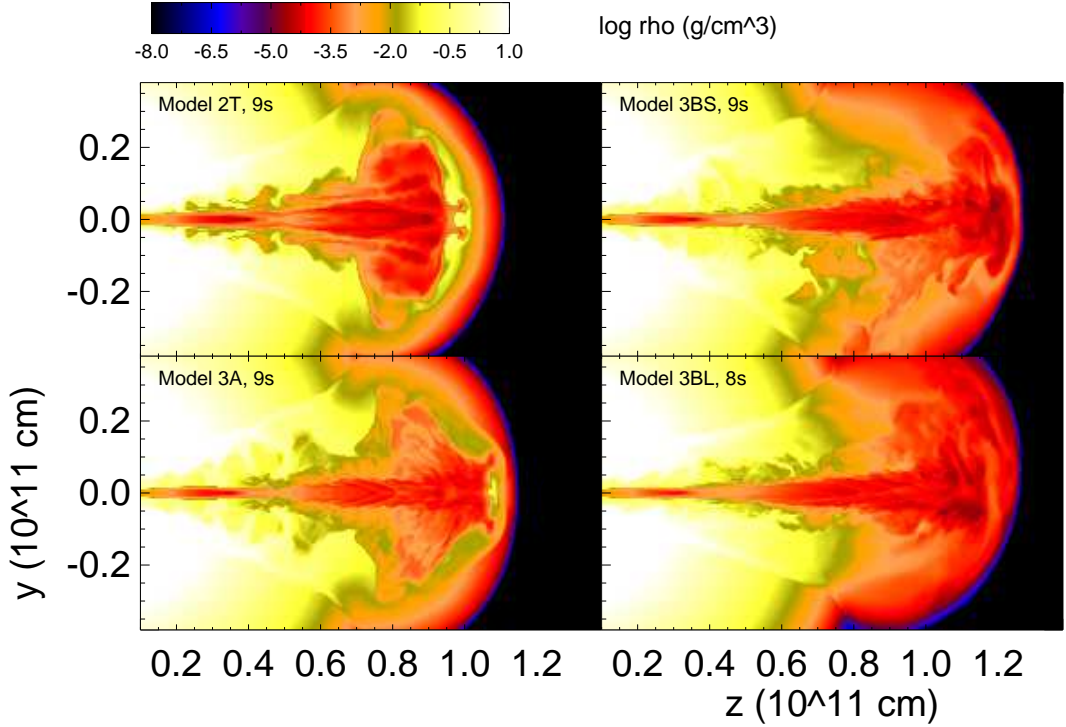}
\caption{A comparison of two- and three-dimensional results at break
out. Slices of the density distribution are shown along the polar
axis.  In the case of the three-dimensional models, the slice shown is
along the $x = 0$ plane.  Model 2T and 3A had the same jet parameters
and effective zoning ($r$ and $z$ in 2T, $y$ and $z$ in 3A) and can
thus be compared. Qualitatively the results at $9\,\Sec$ are similar,
though Model 3A has evolved just a little bit faster. A dense plug of
matter is visible at the head of the jet in both studies.  Even though
3A is a three-dimensional study, it retains the two-dimensional
symmetry imposed by the jet's initial parameters.  Model 3BS is like
Model 3A, but with slightly asymmetric initial conditions, a 1\%
excess of energy in one half of the jet (see text). The asymmetry is
such that a little more energy was deposited in the top half of the
plane displayed than in the bottom half. The result is similar to
Model 3A but now the strict two-dimensional symmetry is broken. The
top of the jet looks quite different from the bottom. Model 3BL
carries the symmetry breaking further with a 10\% imbalance in energy
at the base of the jet. The density of the ``plug'' is greatly
diminished in these asymmetric jets. \label{fig:3dbo} }
\end{figure}

As expected, Model 3A closely resembles Model 2T
(Fig.~\ref{fig:3dbo}). The initial stellar model, the zoning (Tables
\ref{tab:2dgrid} and \ref{tab:3dgrid}), and the jet parameters (Tables
\ref{tab:mod2} and \ref{tab:mod3}) were all the same. In particular,
the initial conditions for the jet have cylindrical symmetry in both
cases. It is gratifying that the answer is insensitive to the
dimensionality of the grid, which in Model 2T, was two-dimensional
cylindrical and, in Model 3A, was three-dimensional Cartesian. This
implies that the bulk properties of jets calculated in Paper 1 -
including collimation and modulation - would be the same in 3D.

However, Model 3A, by assuming a jet that initially has perfect
cylindrical symmetry does not fully exercise the 3D code. Aside from
numerical noise, the equations conserve the 2D symmetry of the initial
conditions, even on a 3D grid. Fig.~\ref{fig:3dbo} also compares, at
breakout, the properties of jets that were, initially, {\sl nearly}
identical.  In particular, Model 3BS differed from 3A only in a 1\%
asymmetry in input energy from one side of the jet to the other, yet
the structure of the emergent jet and cocoon is strikingly
different. In Model 3BL with a 10\% asymmetry the difference is even
more striking. The plots show a cross section of the Cartesian grid
along the initial jet axis in the $x$ = 0 plane. Because of the 40
degree offset (\Sect{3ddef}), there is somewhat more energy in the top
half of the figure than the bottom.  Though both jets retain high
Lorentz factors in their cores, similar to Models 2T and 2B, the
cocoon explosion is a little earlier and larger on the top.  More
dramatic is the difference in the high density ``plug'' among Models
3A, 3BS, and 3BL. In the latter two where the 2D symmetry was mildly
broken, the plug has a much lower density and is not prominent.  In 2D
models and symmetric 3D Model 3A, the plug is held by a concave
surface of the highly relativistic jet core.  Because of the imposed
axisymmetry, the plug cannot easily escape and is pushed forward by
the jet beam.  Whereas the story of the plug is different in
asymmetric 3D Models 3BS and 3BL.  In these models, where the 2D
symmetry was initially mildly broken, instabilities will develop.
The forward bow shock is no longer symmetric and even the stellar
material on the axis is pushed sideways.  More importantly, the head of
the highly relativistic jet beam is also asymmetric and does not have
a concave surface to ``hold'' the plug.  A movie of these runs shows
the plug forming, slipping off to the side, then forming again.  The
plug in these models has a much lower density and is not prominent.
The presence or absence of this plug may have important implication
for the production of short-hard gamma-ray bursts in the massive star
models \citep[Paper 1;][]{Wax03}.

\subsection{Stability of the Jet}

Given the ability to model jets in 3D, we undertook a study to test
their survivability against non-radial instabilities.  Three jets were
introduced on the standard grid (\Tab{3dgrid}) with inclinations to
the radial of 3, 5, and 10 degrees (\Tab{mod3}). These jets were made
to precess with a period equal to $2\,\Sec$, i.e., a non-trivial
fraction of the time it takes the jet to traverse the star and break
out. The results are shown in Fig.~\ref{fig:3dprec}.

With an angle of 3 degrees, the jet in Model 3P3 escapes the star with
its relativistic flow at least partly intact. Though
Fig.~\ref{fig:3dprec} shows Lorentz factors of only $\Gamma \sim 10
-20$, the internal energy is still very large and some of these
regions will attain $\Gamma \sim 180$ if they expand freely.  However,
there is some ``baryon-poisoning''. No clear line of sight exists
along a radial line for the most energetic material even in the jet
beam.  Perhaps a jet with intermediate Lorentz factor would emerge. We
have not yet followed these calculations to large radius (as in
Figs.~\ref{fig:2A}, \ref{fig:2B}, and \ref{fig:2C}) because the 3D
grid was not large enough. Larger calculations are planned.

At larger angles Model 3P5 and especially 3P10 show the break up of
the jet.  Because there is no well focused highly relativistic jet
beam, especially in Models 3P5 and 3P10, more baryon mass is mixed
into the jet.  The Lorentz factor of the jet will decrease due to
dissipation.  And it will be very difficult for these jets to make a
common GRB.  Again some sort of hard transient might be expected,
especially after all the internal energy converts and $\Gamma$ rises,
but for Model 3P10 there will be no GRB of the common variety.  In our
simulations, we find that the critical angle for jet precession is
about 3 degrees.  One would expect that the constraint on the angle of
precession will be reduced if the jet bears more power or is powered
longer.

Is two seconds a reasonable period for the jet to precess?  There are
many uncertainties, but the gravitomagnetic precession period of a
black hole surrounded by an accretion disk is estimated by Hartle,
Thorne, \& Price (1986) to be
$$ T_{\mathrm{GM}} = 900 (M_{\mathrm{disk}}/0.001\,\Msun)^{-1}
                       (M_{\mathrm{bh}}/3\,\Msun)^{-1/2}
                       (r/300\,\km)^{2.5}\,\Sec, $$ 
where $M_{\mathrm{disk}}$ is the mass of the accretion disk,
$M_{\mathrm{bh}}$ is the mass of the hole and $r$ is the radius of the
disk.  $r = 300\,\km$ is a reasonable value for the radius of the
disk, which depends on the distribution of angular momentum of the
star (MacFadyen \& Woosley 1999).  The dependence of the mass in the
disk on the alpha-viscosity has been explored in both numerical
simulations (MacFadyen \& Woosley 1999) and analytic calculations
(Popham, Woosley \& Fryer 1999).  For currently favored value of the
alpha-viscosity, $\sim 0.1$, the disk mass is about $0.001\,\Msun$.
For the disk mass to approach $1\,\Msun$ the viscosity would need to
be $\alpha < 0.001$.  Given the low mass of the accretion disk and the
radius of the disk, $\sim 300\,\km$, precession periods as small as
two seconds are unlikely, but the black hole may be kicked or
instabilities deeper in the region of the star not modeled here could
give the jet some non-radial component.  If so, Fig.~\ref{fig:3dprec},
shows an alternate way in which baryon-loaded jets could give rise to
less relativistic mass ejected and slower moving jets far away from
the star($10 \lesssim \Gamma \lesssim 100$).  Softer transients like
XRFs and sub-luminous GRBs like GRB 980425 could result.

\section{Discussion}
\lSect{disc}

Our calculations show, so long as the orientation of the jet does
not waiver on time scales of order seconds by angles greater than 3
degrees, that a relativistic jet can traverse a Wolf-Rayet star while
retaining sufficient energy and Lorentz factor to make a GRB. This
conclusion is robust in three dimensions as well as two.

\begin{figure}  
\plotone{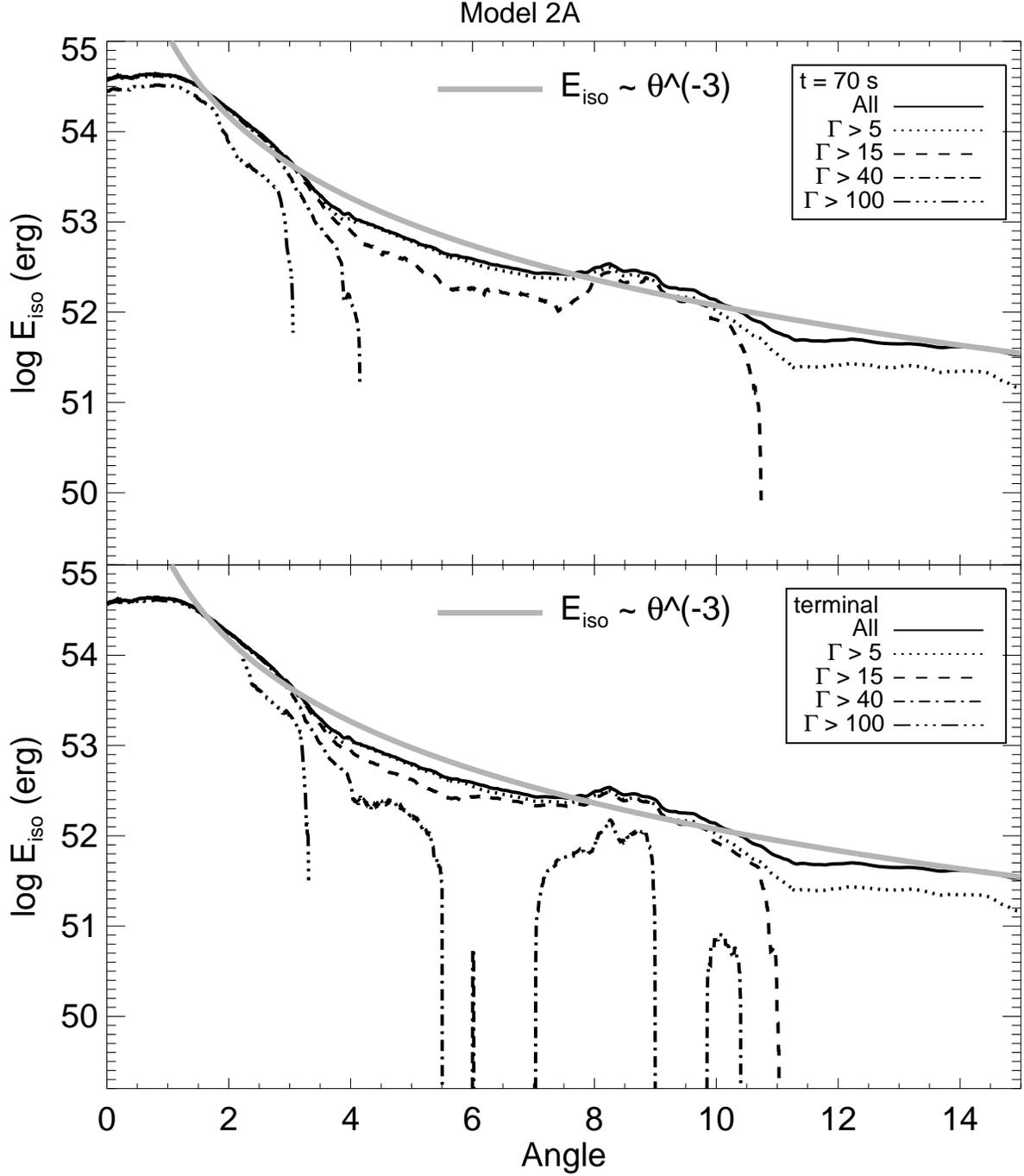}
\caption{Equivalent isotropic energy for Model 2A. As defined in the
text, the equivalent energy to an isotropic explosion inferred by a
viewer at angle $\theta$ is plotted for various Lorentz factors. The
line gives the energy contained in matter with $\Gamma$ greater than
the indicated value moving at a given angle.  The top panel shows this
quantity at 70 s; the bottom panel shows the estimated value much
later, when all internal energy has converted to kinetic.  The light
gray line is a simple power-law fit, $E_{\mathrm{iso}} = 1.5 \times
10^{54} \times (\theta/2\degr)^{-3}\,\erg$. \label{fig:eiso2a}}
\end{figure}

\begin{figure}  
\plotone{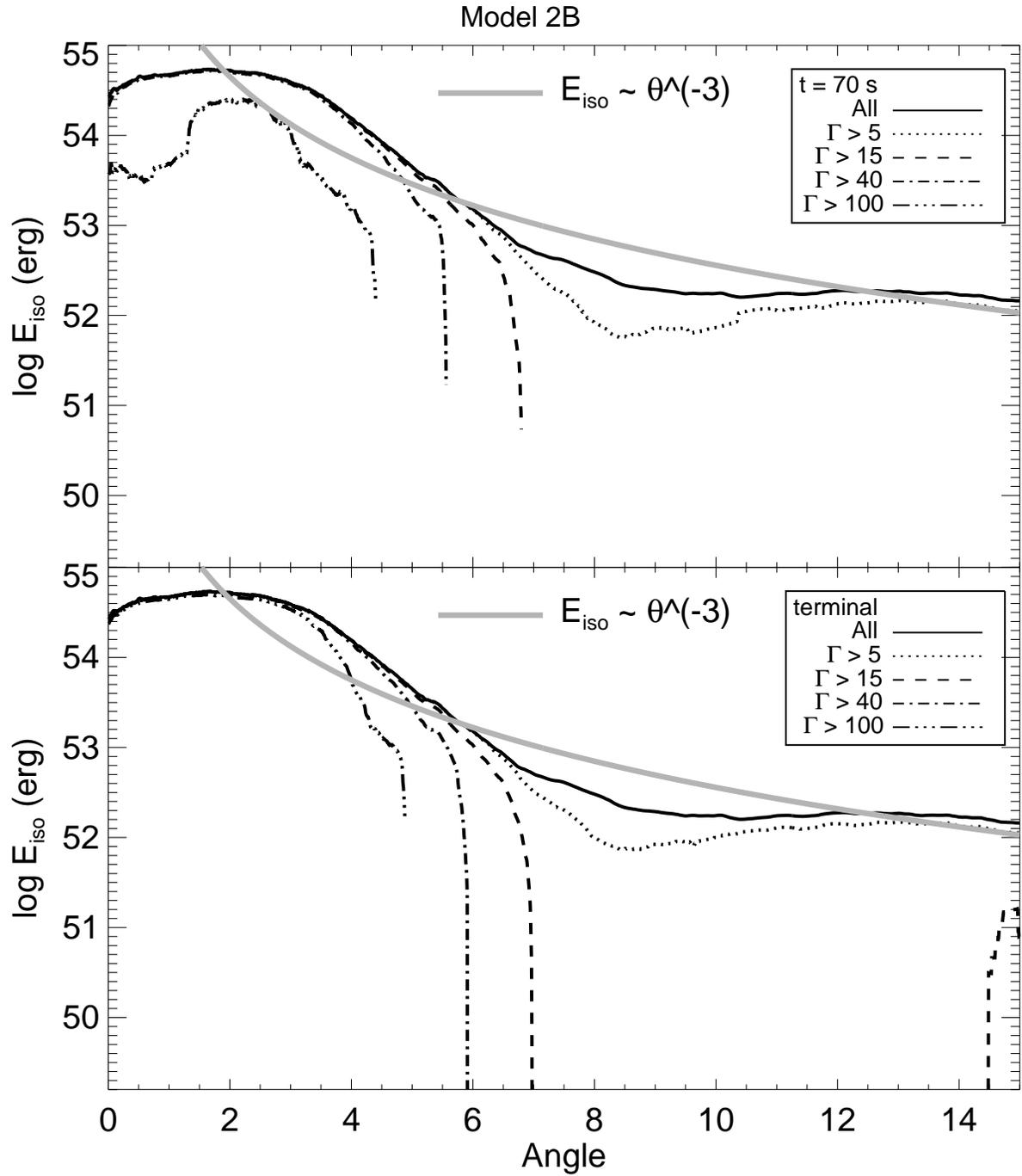}
\caption{Equivalent isotropic energy for Model 2B. The light
gray line is a simple power-law fit, $E_{\mathrm{iso}} = 4.5 \times
10^{54} \times (\theta/2\degr)^{-3}\,\erg$.  See also
Fig.~\ref{fig:eiso2a}. \label{fig:eiso2b}}
\end{figure}

\begin{figure}  
\plotone{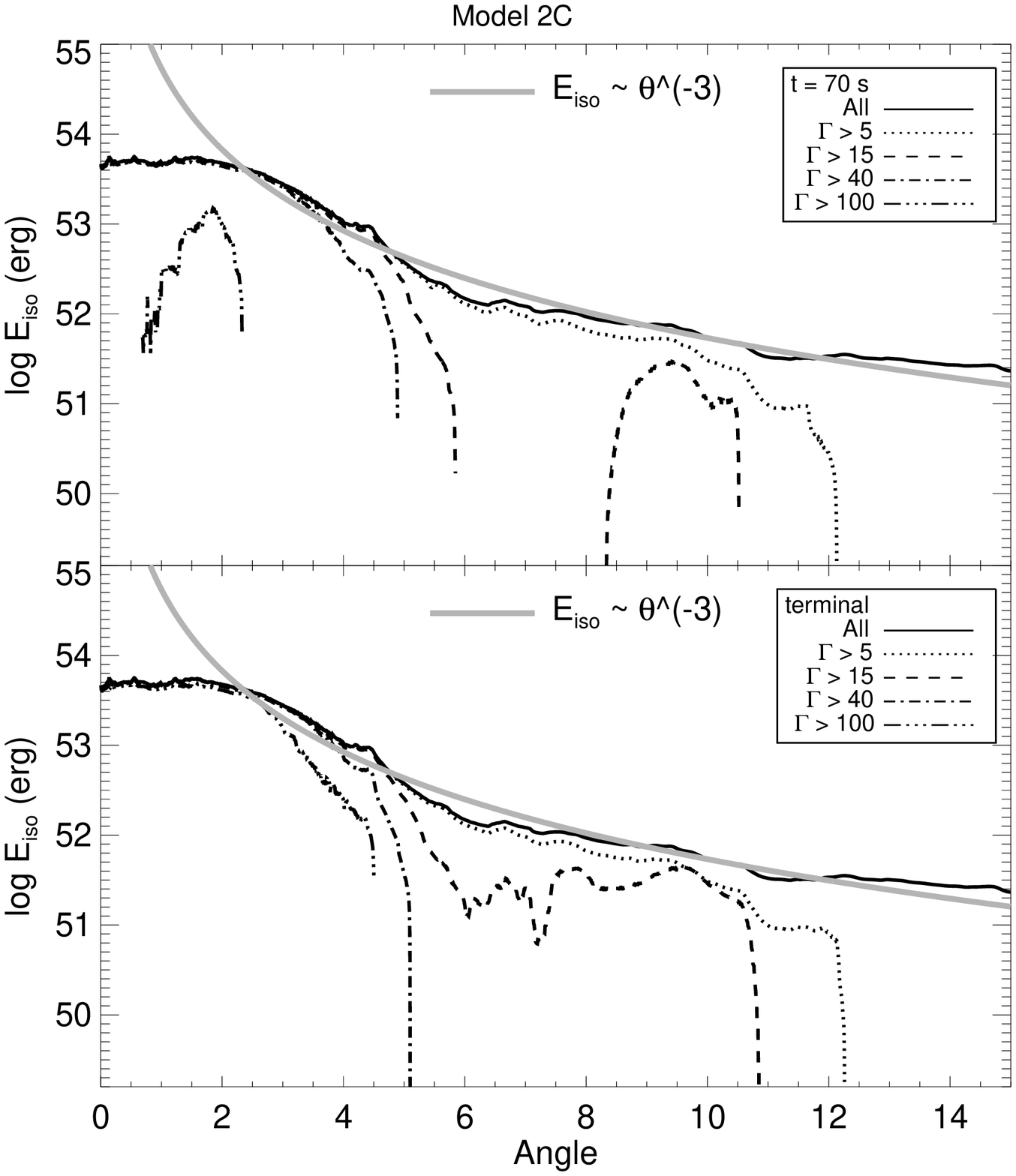}
\caption{Equivalent isotropic energy for Model 2C. The light
gray line is a simple power-law fit, $E_{\mathrm{iso}} = 6.8 \times
10^{53} \times (\theta/2\degr)^{-3}\,\erg$.  See also
Fig.~\ref{fig:eiso2a}. The solid angle between 8 and 10 degrees is 4
times that inside of 3 degrees.  That is, the observers are more likely
to be off-axis than on-axis. \label{fig:eiso2c}}
\end{figure}

As it breaks out, the jet is surrounded by a cocoon of
mildly-relativistic, energy-laden matter. As internal energy converts
to kinetic, matter with $10^{51}$ to $10^{52}\,\erg$ of equivalent
isotropic energy moving with Lorentz factor $\Gamma \gtrsim 20$ is
ejected to angles about three times greater than the GRB. That is,
whatever transient the cocoon gives rise to will be an order of
magnitude more frequently observable in the universe, but two orders
of magnitude less energetic than a GRB. Weaker transients can of
course be obtained at still larger angles and there is considerable
diversity in the models (Figs.~\ref{fig:eiso2a}, \ref{fig:eiso2b}, and
\ref{fig:eiso2c}).

The isotropic equivalent rest mass to $10^{51}$ - $10^{52}\,\erg$ with
$\Gamma \approx 20$ is $3 \times 10^{-5}$ to $3 \times
10^{-4}\,\Msun$. This matter would give up its energy upon
encountering $1/\Gamma$ times its rest mass. For an assumed mass loss
rate for the Wolf-Rayet progenitor of $10^{-5}\,\Msunyr$ and speed,
$1000\,\kms$, the energy will be radiated at about $3 \times 10^{14}$
to $3 \times 10^{15}\,\cm$. Emission from this deceleration will have
a duration $\sim R/(2 \Gamma^2 c) \sim$ 10 - $100\,\Sec$. Variations
of a factor of 10 are easily achievable by varying the mass loss rate
or Lorentz factor.

Unlike the central jet, we do not see evidence for large scale
variation in the Lorentz factor at lower latitude and so it would be
hard to make the transient emission by internal shocks. We thus
attribute the emission to external shocks and invoke a variety of
Lorentz factors emitting within our light cone in order to explain the
spread in wavelength.

Might there be observable counterparts to these large angle, low
Lorentz factor explosions? Typically, they have too low a Lorentz
factor to make common GRBs \citep{Gue01}. \citet{Tan01} present
arguments however that, by external shock interaction with the
progenitor wind, a hard transient of some sort should result.

Our model predicts a correlation between $E_{\rm iso}$, the Lorentz
factor, and the angle between the polar axis and the observer
(Fig.~\ref{fig:eiso2a}, \ref{fig:eiso2b} and \ref{fig:eiso2c}).
Roughly speaking, the GRB outflows have a narrow highly relativistic
jet beam and a wide mildly relativistic jet wing.  Recent observations
and afterglow modeling support this non-uniform jet model
\citep{ber03,zhab03,sal03,ros02}.
If, as seems reasonable, the Lorentz factor is, in turn, correlated
with the peak energy observed in the burst one expects a continuum of
high energy transients spanning the range from X-ray afterglows (keV),
to hard X-ray transients (tens of keV), to GRBs (hundreds of
keV). Observations reviewed by \citet{Ama02} show such a correlation
for bursts with $E_{\rm peak}$ from 80 keV to over 1 MeV and Lamb
(private communication) finds that the relation extends to the lower
energies.

In particular, XRFs form a new class of X-ray transients having a
duration of order minutes and properties that in ways resemble GRBs
\citep{Hei01,Kip03}.  XRFs are probably many phenomena \citep{are03}
and it could be that, especially some of the longer ones, have
alternate explanations (see introduction). However, we have felt for
some time \citep[][Paper 1]{Woo99,woo99b,Woo00,Woo01} that many XRFs
are the off-axis emissions of GRBs, made in the lower energy wings of
the principal jet. We have not calculated the expected spectra for any
of our models, only Lorentz factors and energies. However, it is
reasonable that matter moving with $\Gamma \sim$ 20 would make a
transient softer than a GRB and harder than a few keV. Larger amounts
of energy are emitted at a larger range of angles for slower but still
relativistic ejecta ($\Gamma = 5$, e.g., Fig.~\ref{fig:eiso2a}).

If our model is valid, XRFs and GRBs should be continuous classes of
the same basic phenomenon sharing many properties. They should have a
similar spatial distribution to GRBs because they are essentially the
same sources. However, because they are much less luminous, their log
N-log S distribution would not exhibit the same roll over attributed
in GRBs to seeing the ``edge'' of the distribution (e.g., Fishman \&
Meegan 1995).  Their median
redshift should be considerably smaller, certainly less than 1. This
is consistent with the low redshifts inferred for the host galaxies of
two XRFs by \citet{Blo03}. XRFs may, most frequently, be seen in
isolation and will be characterized by softer spectra, but there would
also be an underlying XRF in every GRB since the emission of the
mildly relativistic cocoon material is beamed to a larger angle that
includes the poles. In some cases these XRFs might be seen as
precursors or extended hard X-ray emission following a common
GRB. They should be associated with supernovae. Indeed XRFs may more
frequently serve as guideposts to jet-powered supernovae than GRBs,
especially the nearby ones.

Though considerable variation is expected, our calculations (e.g.,
Fig.~\ref{fig:eiso2a}) suggest that XRFs are typically visible at
angles about three times greater than GRBs and hence to ten times the
solid angle. However, their energy, a few percent of GRBs, implies
that a flux-limited sample could observe GRBs out to roughly ten times
farther, implying that XRFs would be about 1\% as frequent in the
sample. The actual value is detector sensitive, but may be larger.

\begin{figure}  
\plotone{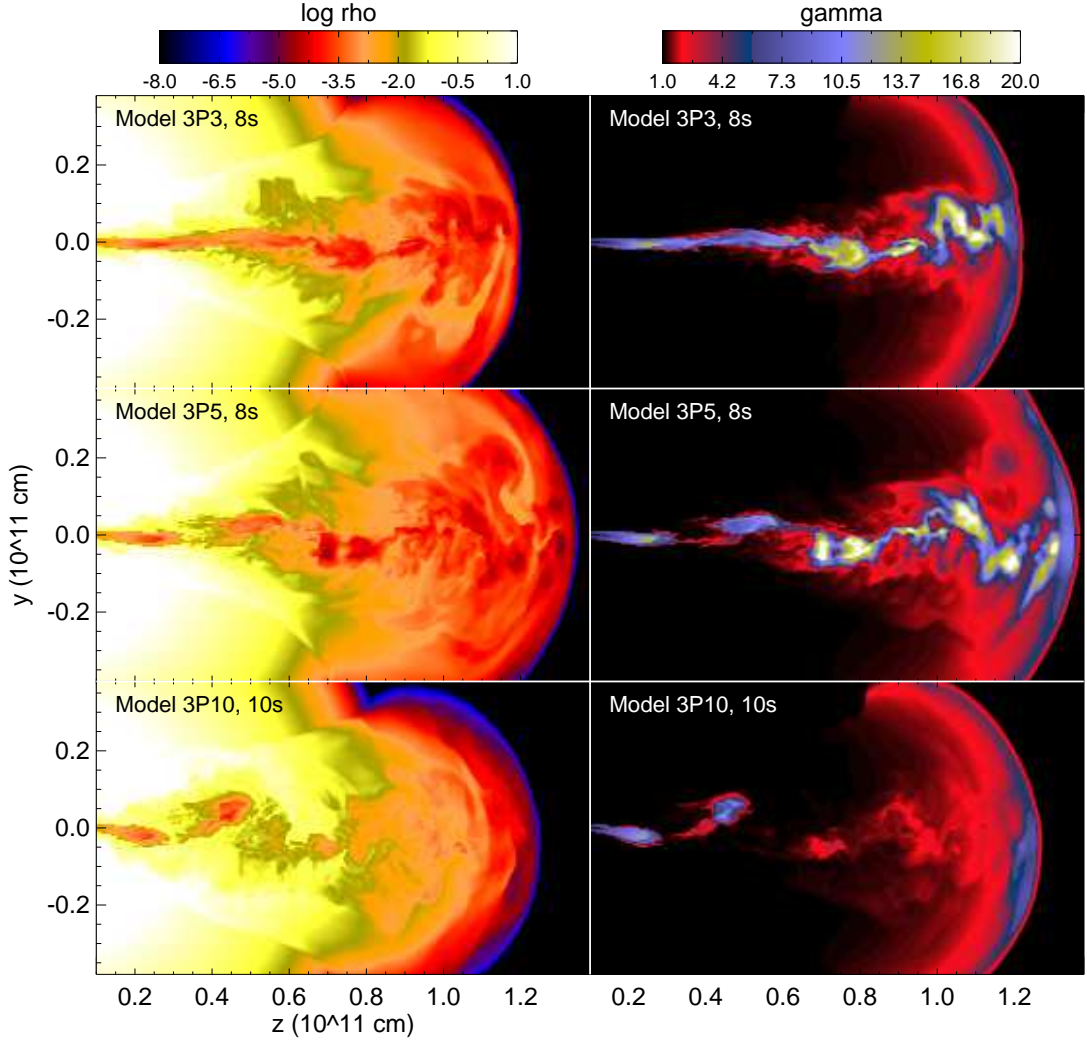}
\caption{Precessing jets - angle sensitivity study. Slices of the
precessing jet models defined in \Tab{mod3} and \Sect{3ddef}
corresponding to the $x = 0$ plane are shown just after break out. The
three jets had an inclination with respect to the radial of 3, 5, and
10 degrees, respectively, for models 3P3, 3P5, and 3P10 and the
initial jet precessed on a cone with this half angle with a period of
$2\,\Sec$.  For Model 3P3, the jet still emerges relatively intact
(though one would want to follow the evolution further before
concluding a GRB is still produced). The jets in Models 3P5 and
especially 3P10 dissipate their energy before escaping and are
unlikely to give terminal Lorentz factors as high as 200 (though they
may still produce hard X-ray flashes). \label{fig:3dprec} }
\end{figure}

One should also keep in mind the possibility that XRFs do {\sl not}
accompany GRBs, in any direction, but are a result of baryon loading
of the central jet. Three-dimensional instabilities
(Fig.~\ref{fig:3dprec}) may play an important role in this and are
being explored.

\acknowledgments

This research has been supported by NASA (NAG5-8128, NAG5-12036, and
MIT-292701) and the DOE Program for Scientific Discovery through
Advanced Computing (SciDAC; DE-FC02-01ER41176).  This research used
resources of the National Energy Research Scientific Computing Center,
which is supported by the Office of Science of the U.S. Department of
Energy under Contract No. DE-AC03-76SF00098.  AH was supported, in
part, by the Department of Energy under grant B341495 to the Center
for Astrophysical Thermonuclear Flashes at the University of Chicago,
and a Fermi Fellowship at the University of Chicago.  We appreciate
the helpful comments of the referee.


\clearpage
\begin{deluxetable}{cccc}
\tablecaption{Properties\tablenotemark{a} \ \ of the Progenitor Model \label{tab:proj}}
\tablewidth{0pt}
\tablehead{
\colhead{} &
\colhead{m} &
\colhead{r} &
\colhead{J} 
\\
\colhead{} &
\colhead{($\Msun$)} &
\colhead{($\cm$)} &
\colhead{($\erg\ $s)}
}
\startdata
iron core    &  1.95 & $2.58\E{ 8}$ & $1.23\E{50}$ \\
Si core      &  2.61 & $4.99\E{ 8}$ & $2.35\E{50}$ \\
Ne/Mg/O core &  2.95 & $7.06\E{ 8}$ & $2.68\E{50}$ \\
C/O core     &  8.56 & $5.16\E{ 9}$ & $2.24\E{51}$ \\
star / He core & 15.00 & $8.80\E{10}$ & $1.00\E{52}$ 
\enddata
\tablenotetext{a}{Enclosed mass, $m$, radius, $r$, and enclosed angular
momentum, $j$, of the progenitor model at core collapse on the outer
boundaries of the indicated cores.}
\end{deluxetable}

\clearpage
\begin{deluxetable}{ccccccccc}
\tablecaption{Parameters of 2D-Models \label{tab:mod2}}
\tablewidth{0pt}
\tablehead{
\colhead{} &
\colhead{$\dot{E}$\tablenotemark{a}} &
\colhead{} &
\colhead{} &
\colhead{$z_0$\tablenotemark{d}} &
\colhead{$T_1$\tablenotemark{e}} &
\colhead{$T_2$\tablenotemark{f}} &
\colhead{$E_{\mathrm{tot}}$\tablenotemark{g}} &
\colhead{} \\
\colhead{Model} &
\colhead{($10^{50}\,\ergs$)} &
\colhead{$\Gamma_{0}$\tablenotemark{b}} &
\colhead{$f_{0}$\tablenotemark{c}} &
\colhead{($10^{10}\,\cm$)} &
\colhead{($\Sec$)} &
\colhead{($\Sec$)} &
\colhead{($10^{51}\,\erg$)} &
\colhead{Grid\tablenotemark{h}}
}
\startdata
 2A & 1.0 & 10 & 20 & 1.0 & 20 & 10 &  5.0 & Normal \\
 2B & 3.0 &  5 & 40 & 1.0 & 20 & 10 & 15.0 & Normal \\
 2C & 0.5 &  5 & 40 & 1.0 & 20 & 10 &  2.5 & Large \\
 2T & 3.0 &  5 & 40 & 1.0 & 20 & 10 & 15.0 & Coarse 
\enddata
\tablenotetext{a}{Energy deposition rate per jet}
\tablenotetext{b}{Initial Lorentz factor}
\tablenotetext{c}{Initial ratio of the total energy (excluding the
  rest mass) to the kinetic energy}
\tablenotetext{d}{\textnormal{Low-$z$ boundary, where the jet was injected}} 
\tablenotetext{e}{During this period, the jet power remained constant.}
\tablenotetext{f}{During this period, the jet power was turned down linearly.}
\tablenotetext{g}{Total energy injected during the calculation}
\tablenotetext{h}{See Table~\ref{tab:2dgrid} for details}

\end{deluxetable}

\clearpage
\begin{deluxetable}{cccccc}
\tablecaption{Energetics of 2D-Models\tablenotemark{a} \label{tab:energy}} 
\tablewidth{0pt}
\tablehead{ 
\colhead{Model} & 
\colhead{Energy input} & 
\colhead{Total on grid\tablenotemark{b}} &
\colhead{$v > 0.5 \, c$} & 
\colhead{$\Gamma > 2$} &
\colhead{$\Gamma > 10$} 
\\ 
\colhead{} &
\colhead{(10$^{51}$ erg)} & 
\colhead{(10$^{51}$ erg)} & 
\colhead{(10$^{51}$ erg)} & 
\colhead{(10$^{51}$ erg)} & 
\colhead{(10$^{51}$ erg)}  
} 
\startdata
 2A & 5.0  & 4.66  & 3.64  & 3.56 & 3.32  \\
 2B & 15   & 14.36 & 12.46 & 12.22 & 11.16 \\
 2C & 2.5  & 1.93  & 1.20  & 1.13  & 0.92 
\enddata
\tablenotetext{a}{Energies evaluated for the entire star, both jets at
40 s.}
\tablenotetext{b}{The calculations conserve energy to an accuracy of
better than 0.1\%.  However, some energy has left the computational grid.}

\end{deluxetable}

\clearpage
\begin{deluxetable}{ccccccccc}
\tablecaption{Zoning in 2D-Models ($10^{10}\,\cm$) \label{tab:2dgrid}} 
\tablewidth{0pt}
\tablehead{ 
\colhead{Model} & 
\colhead{$r_1$\tablenotemark{a}} & 
\colhead{$r_2$} &
\colhead{$r_3$} & 
\colhead{$z_1$} &
\colhead{$z_2$} & 
\colhead{$z_3$} &
\colhead{$r$} & 
\colhead{$z$} 
\\ 
\colhead{} &
\colhead{$\Delta r_1$} & 
\colhead{$\Delta r_2$} & 
\colhead{$\Delta r_3$} & 
\colhead{$\Delta z_1$} & 
\colhead{$\Delta z_2$} & 
\colhead{$\Delta z_3$} &  
\colhead{zones} & 
\colhead{zones} 
} 
\startdata
 2A & 10.0 & 20.0 & 60.0 & 10.0 & 20.0 & 200 & 1500 & 2275 \\
    & 0.01 & 0.04 & 0.16 & 0.01 & 0.04 & 0.16 & & \\
 2B & 10.0 & 20.0 & 60.0 & 10.0 & 20.0 & 200 & 1500 & 2275 \\
    & 0.01 & 0.04 & 0.16 & 0.01 & 0.04 & 0.16 & & \\
 2C & 10.0 & 20.0 &200.0 & 10.0 & 20.0 & 200 & 2375 & 2275 \\
    & 0.01 & 0.04 & 0.16 & 0.01 & 0.04 & 0.16 & & \\
 2T & 0.64 & 3.84 & - & 13.8 & - & - & 256 & 512 \\
    & 0.01 & 0.05 & - & 0.025 & - & - & & 
\enddata
\tablenotetext{a}{The first row for each model gives the upper bound
of distance ($r$, or $z$) for which the zoning indicated in the second
row was employed. All are measured in units of $10^{10}\,\cm$. $r$
starts at 0; $z$ starts at $1.0 \times 10^{10}\,\cm$. The zoning in
Models 2A and 2B was identical.}
\end{deluxetable}

\clearpage
\begin{deluxetable}{cccccc}
\tablecaption{Energetics of 2D-Models 2B and 2T
                 \tablenotemark{a} \label{tab:2bt}} 
\tablewidth{0pt}
\tablehead{ 
\colhead{Model} & 
\colhead{Energy input} & 
\colhead{Total energy} &
\colhead{$v > 0.5 \, c$} & 
\colhead{$\Gamma > 2$} &
\colhead{$\Gamma > 10$} 
\\ 
\colhead{} &
\colhead{(10$^{51}$ erg)} & 
\colhead{(10$^{51}$ erg)} & 
\colhead{(10$^{51}$ erg)} & 
\colhead{(10$^{51}$ erg)} & 
\colhead{(10$^{51}$ erg)}  
} 
\startdata
 2B & 3  & 2.46  & 1.44  & 1.31 & 0.51 \\
 2T & 2.85 & 2.42 & 1.50 & 1.13 & 0.52 
\enddata
\tablenotetext{a}{Energies for Model 2T are evaluated for the entire
                 computational grid at $9.5\,\Sec$, whereas, energies
                 for Model 2B are evaluated for only part of the
                 whole grid at $10\,\Sec$.  Note that the grid of
                 Model 2B is larger 
                 than that of Model 2T (Table~\ref{tab:2dgrid}).  We
                 only consider part of the grid for Model 2B so that
                 the same physical region in the two models are
                 discussed.  The bow shocks of the jets are at similar
                 radius for the two models at the chosen moment
                 (Fig.~\ref{fig:2bt}).
		 Also note that we only consider one hemisphere.} 

\end{deluxetable}

\clearpage
\begin{deluxetable}{ccccccc}
\tablecaption{Parameters of 3D-Models \label{tab:mod3}}
\tablewidth{0pt}
\tablehead{
\colhead{} &
\colhead{$\dot{E}$} &
\colhead{} &
\colhead{} &
\colhead{$z_0$} &
\colhead{$\theta_p$\tablenotemark{a}} &
\colhead{Period\tablenotemark{b}} \\
\colhead{Model} &
\colhead{($10^{50}\,\ergs$)} &
\colhead{$\Gamma_{0}$} &
\colhead{$f_{0}$} &
\colhead{($10^{10}\,\cm$)} &
\colhead{(degrees)} &
\colhead{($\Sec$)}  
}
\startdata
 3A     & 3.0 & 5 & 40 & 1.0 & 0  & -  \\
 3BL\tablenotemark{c}    & 3.0 & 5 & 40 & 1.0 & 0  & -  \\
 3BS\tablenotemark{c}    & 3.0 & 5 & 40 & 1.0 & 0  & -  \\
 3P3 & 3.0 & 5 & 40 & 1.0 & 3  & 2  \\
 3P5 & 3.0 & 5 & 40 & 1.0 & 5  & 2  \\
 3P10 & 3.0 & 5 & 40 & 1.0 & 10 & 2 
\enddata
\tablenotetext{a}{Angle between the axis of the injected jet and the $z$-axis}
\tablenotetext{b}{Period of precession in Models 3P3, 3P5, and 3P10}
\tablenotetext{c}{Models 3BL and 3BS were like Model 3A (and Model 2B)
but included an asymmetric energy input at the base of 10\% in Model
3BL and 1\% in Model 3BS.}

\end{deluxetable}

\clearpage
\begin{deluxetable}{cccccccc}
\tablecaption{Zoning in 3D-Models ($10^{10}\,\cm$) \label{tab:3dgrid}} 
\tablewidth{0pt}
\tablehead{ 
\colhead{$x_1$\tablenotemark{a}} & 
\colhead{$x_2$} &
\colhead{$y_1$} & 
\colhead{$y_2$} &
\colhead{$z_1$} & 
\colhead{$x$} & 
\colhead{$y$} & 
\colhead{$z$} 
\\ 
\colhead{$\Delta x_1$} & 
\colhead{$\Delta x_2$} & 
\colhead{$\Delta y_1$} & 
\colhead{$\Delta y_2$} & 
\colhead{$\Delta z_1$} &   
\colhead{zones} & 
\colhead{zones} & 
\colhead{zones} 
} 
\startdata
 0.64 & 3.84 & 0.64 & 3.84 & 13.8  & 256 & 256 & 512 \\
 0.01 & 0.05 & 0.01 & 0.05 & 0.025 &     &     &      
\enddata
\tablenotetext{a}{The first row for each model gives the {\sl{absolute}}
value of the upper bound of distance ($x$, $y$, or $z$) for which the
zoning indicated in the second row was employed. All are measured in
units of $10^{10}\,\cm$. $|x|$ and $|y|$ start at 0; $z$ starts at
$1.0 \times 10^{10}\,\cm$. All three-dimensional studies used this
same zoning.}

\end{deluxetable}

\clearpage

\clearpage

\clearpage

\clearpage

\clearpage

\clearpage

\clearpage

\clearpage

\clearpage

\clearpage

\clearpage

\clearpage

\end{document}